\documentclass[fleqn,twoside]{article}
\usepackage[headings]{espcrc2}

\readRCS
$Id: espcrc2.tex,v 1.2 2004/02/24 11:22:11 spepping Exp $
\usepackage{graphicx}
\usepackage[figuresright]{rotating}

\newcommand\as{\alpha_s}

\def\JHEP{{\it Jour.~High~Energy~Phys.~}}
\def\vol#1{{\bf #1}}\def\vyp#1#2#3{\vol{#1} (#2) #3}

\catcode`@=11 %This allows us to modify plain macros
\def\slash#1{\mathord{\mathpalette\c@ncel#1}}
 \def\c@ncel#1#2{\ooalign{$\hfil#1\mkern1mu/\hfil$\crcr$#1#2$}}
\def\lsim{\mathrel{\mathpalette\@versim<}}
\def\gsim{\mathrel{\mathpalette\@versim>}}
 \def\@versim#1#2{\lower0.2ex\vbox{\baselineskip\z@skip\lineskip\z@skip
       \lineskiplimit\z@\ialign{$\m@th#1\hfil##$\crcr#2\crcr\sim\crcr}}}

\newcommand{\AmS}{{\protect\the\textfont2
  A\kern-.1667em\lower.5ex\hbox{M}\kern-.125emS}}

\title{Can we trust small $x$ resummation?}

\author{Stefano Forte,\address[]{Dipartimento di  Fisica, Universit\`a di
Milano and  INFN, Sezione di Milano,\\ Via Celoria 16, I-20133 Milan, Italy}
Guido~Altarelli,\address{Dipartimento di Fisica ``E.Amaldi'', 
Universit\`a Roma Tre and INFN, Sezione di Roma Tre\\
Via della Vasca Navale 84, I--00146 Roma, Italy, \\ 
 CERN, Department of Physics, Theory Division\\ CH-1211 Gen\`eve 23,
 Switzerland} 
Richard D.~Ball \address{School of Physics, University of
Edinburgh\\ Edinburgh EH9 3JZ, Scotland}} 
\begin{document}

\begin{abstract}
We review the current status of small $x$ resummation of evolution of
parton distributions and of deep--inelastic coefficient functions. We
show that the resummed perturbative expansion is stable, robust upon
different treatments of subleading terms, and that it matches smoothly to
the unresummed perturbative expansions, with corrections which are of the
same order as the typical NNLO ones in the HERA kinematic region. We
discuss different approaches to small $x$ resummation: we show that
the ambiguities in the resummation procedure are small, provided all
parametrically enhanced terms are included in the resummation and
properly matched.
\vspace{1pc}
\end{abstract}

\maketitle
%\footnote{Talk presented by
%    S.F. at the Ringber Workshop 2008}
%\begin{flushright}
%IFUM-935-FT
%\end{flushright}

\section{The need for small $x$ resummation}
\label{sec:nnedsx}
The so-called small $x$ regime of QCD is the kinematical region in
which hard scattering processes happen at a center-of-mass
energy which is much larger than the characteristic hard scale
of the process. An understanding of strong interactions in this region
is therefore necessary to do physics at high--energy colliders. 
In this sense, HERA was
the first small $x$ machine, and LHC is going to be even more of a
small $x$ accelerator. 

\begin{figure}[htb]
\begin{center}
\vspace{-1.cm}
\includegraphics[width=.75\linewidth]{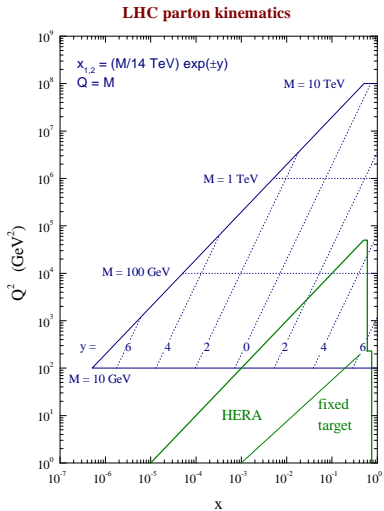} 
\end{center}
\vskip-1.2cm
\caption{Comparison of the HERA and LHC kinematical regions (from
  Ref.~\cite{dittmar}).}
\label{fig:kinreg}
\vskip-.5cm
\end{figure}
In deep--inelastic lepton--hadron interactions, the scale is set by
the virtuality of the photon $Q^2$, and $x=\frac{Q^2}{2 p.q}=
\frac{Q^2}{s}(1+O(x))$, where $s$ is the center--of--mass energy of the
virtual photon--hadron collision, and the distribution of partons which carry
a fraction $x$ of the incoming nucleon energy is probed.
In hadron--hadron interactions, the
scale is set by the invariant mass $M^2$ of the final state, and $x=x_1
x_2$,
with $x_i= \frac{M}{\sqrt{s}}e ^{\pm y}$, $s$ the center--of--mass
energy and $y$ the rapidity of the
hadron--hadron collision. Here, the distribution of partons which
carry fractions $x_i$ of the two incoming nucleon energies are
probed. This means that typical $x$ values probed at the LHC in the
central rapidity region are almost two orders of magnitude smaller
than $x$ values probed at HERA at the same scale. Hence, small $x$
corrections start being relevant even for 
a final state with a characteristic electroweak scale
$M\sim 100$~GeV (see Fig.~\ref{fig:kinreg}).

\begin{figure}[htb]
\begin{center}
\includegraphics[width=.7\linewidth]{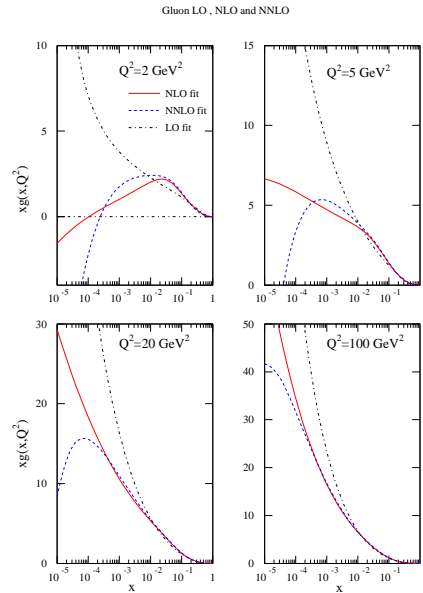} 
\end{center}
\vskip-1.2cm
\caption{Comparison of the LO, NLO and NNLO gluon distributions in the
  MSTW08 parton fit (from
  Ref.~\cite{thorne}).}
\label{fig:instability}
\vskip-.5cm
\end{figure}
As is by now well known, perturbative corrections become large at
small $x$. Due to the accidental vanishing of some
coefficients, the leading large corrections cannot be seen in NLO and
NNLO splitting functions; however, the first subleading correction can
already be seen in the NNLO splitting functions which have been
computed recently, as well as in NNLO coefficient functions: they are
large enough to make recent NNLO parton fits unstable at small
$x$ (see Fig.~\ref{fig:instability}). 

This suggests dramatic effects from yet higher
orders, so the success of NLO perturbation theory at HERA, as
demonstrated by the scaling laws it predicts~\cite{DAS,FC}, has been
for a long time very hard to explain.
In the last several years this situation has been clarified, mostly
thanks to the effort of two groups (ABF~\cite{sxap}-\cite{sxsym} and
CCSS~\cite{salam}-\cite{ciafdip}), which have presented a full
resummation of evolution equations in the gluon sector, thereby
showing that, once all relevant large terms are included, the effect of
the resummation of terms which are enhanced at small $x$ is
perceptible but moderate --- comparable in size to typical NNLO fixed
order GLAP corrections in the HERA region. A detailed comparison of
the ABF and CCSS approach in the pure gluon ($n_f=0$) case was
presented in Ref.~\cite{dittmar}, where excellent agreement was found.

This approach has now been generalized by ABF~\cite{symphen} 
to a full resummation
including quarks, and including the resummation of deep-inelastic
coefficient functions, so that resummed 
expressions for deep-inelastic structure
functions can be obtained. Progress towards the inclusion of quarks
has also been made by the CCSS group~\cite{ciafquarks}, though full
results are not yet available. Meanwhile, an alternative somewhat
simplified approach has also been developed~\cite{T1}, and used to
perform a fit to deep-inelastic scattering data~\cite{TW}. In this
approach, the factorization scheme is not defined in a fully
consistent way at the resummed level, and also
some contributions to the resummation are either not
included, or treated by means of truncated expansions. A comparison of
resummed predictions for deep-inelastic structure functions obtained
in the TW and ABF schemes for resummation is presented in
Ref.~\cite{heralhc08}. 

Here, we shall review the main ingredients that go into
small $x$ resummation, with the aim of understanding the
impact of various contribution on the final result and its
stability upon higher order perturbative corrections and upon the inclusion of
various subleading contributions: we shall take the ABF approach as a
baseline, and discuss the impact of alternative options.
We shall show that small $x$ resummation is very constrained by various
requirements, which include momentum conservation, the
inclusion of collinear contribution and matching to GLAP evolution,
and consistency with the renormalization group. Once all these
requirements are met, further subleading ambiguities are quite small;
however, if some of the corresponding terms are missed out, effects are not
negligible as we shall see. 
The resummed corrections thus obtained  are perturbatively
stable, as demonstrated by the fact that renormalization and
factorization scale dependence are moderate, and decrease with
increasing perturbative order as they ought to. The typical effect of
the resummation in the HERA and LHC regions is comparable to that of
NNLO corrections, but with the opposite sign.

In the next section, we shall review the ingredients which are
necessary in order to perform the resummation in the gluon sector. In
the subsequent section, we shall summarize the generic features of the
resummed results. In the last section, we shall discuss how quarks and
deep-inelastic coefficient functions may be included in the resummation, and
discuss resummed results for deep-inelastic physical observables.
\section{The three ingredients of stable resummation}
\label{sec:threeing}
In this section, we discuss the resummation of evolution equations when $n_f=0$. In this case, there is a single parton
distribution, the gluon distribution $G(\xi,t)$, with
$\xi\equiv\ln\frac{1}{x}$, $t\equiv\ln\frac{Q^2}{\Lambda^2}$. It is
convenient to define the Mellin transforms
\begin{eqnarray}
G(N,t)&\equiv&\int^{\infty}_{0}\! d\xi\, e^{-N\xi}~G(\xi,t)\nonumber\\
G(\xi,M)&\equiv&\int^{\infty}_{-\infty}\! dt\, e^{-Mt}~G(\xi,t)\label{mellins}
\end{eqnarray}
\subsection{Double--leading expansion}
\begin{figure}[htb]
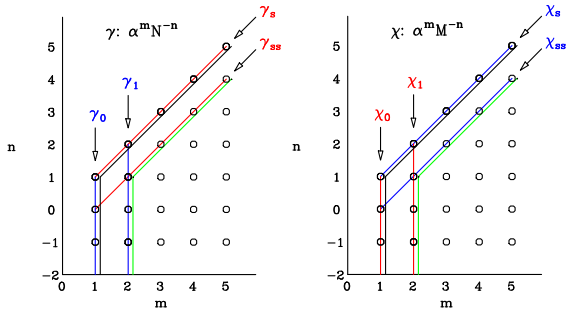

\begin{center}
\includegraphics[width=.49\linewidth]{abfcrosgamdl.ps} 
\includegraphics[width=.49\linewidth]{abfcroschidl.ps} 
\end{center}
\vskip-1.2cm
\caption{Double leading expansion of the GLAP anomalous dimension $\gamma$
  (left) and the BFKL kernel $\chi$ (right).}
\label{fig:crosses}
\vskip-.5cm
\end{figure}
The gluon distribution $G(N,t)$ can be expressed in terms
of the gluon distribution at $t=t_0$ by solving the GLAP equation
\begin{equation}
\frac {d}{dt}G(N,t)=\gamma(N,\as)~G(N,t)
\label{glap}
\end{equation}
which at the N$^k$LO sums all terms of order $\alpha_s^n t^{n-k}$, to
all orders in $\alpha_s$.
 The first step of resummation consists of including, to the  N$^k$L
 log level,   all contributions
to the anomalous dimension $\gamma(N,\as)$ of order $\alpha_s^n N^{-(n-k)}$, to
all orders in $\alpha_s$, since they correspond to contributions of
order $\alpha_s^n \ln^{n-k}\frac{1}{x}$ to $G(\xi,t)$.

This inclusion is straightforward at the fixed coupling level, thanks to
the fact that the gluon distribution $G(\xi,M)$ can be expressed in terms
of the gluon distribution at $\xi=\xi_0$ by solving the BFKL equation
\begin{equation}
\frac {d}{d\xi}G(\xi,M)=\chi(M,\as)~G(\xi,M),
\label{bfkl}
\end{equation}
whose kernel $\chi(M,\as)$ is simply the inverse function of 
the GLAP anomalous dimension
$\gamma(N,\as)$~\cite{jaro,sxap}:
\begin{equation}
\chi(\gamma(N,\as),\as) =   N.
\label{dual}
\end{equation}

\begin{figure}[htb]
\begin{center}
\includegraphics[width=\linewidth]{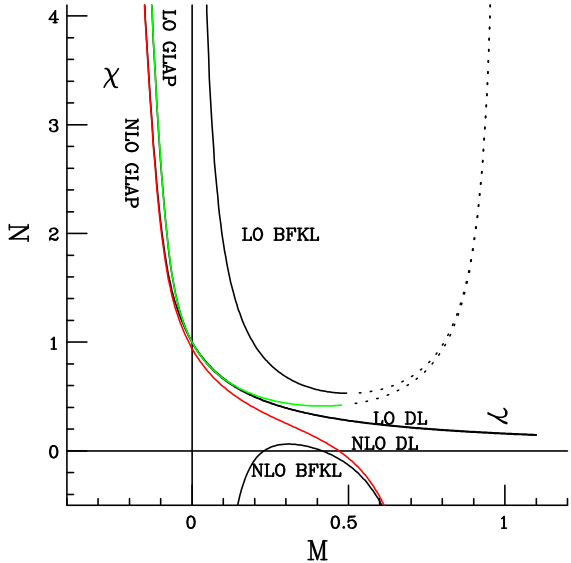} 
\end{center}
\vskip-1.2cm
\caption{The BFKL kernel and its dual GLAP anomalous dimension
  computed at LO and NLO in the BFKL expansion, the GLAP expansion and
  the double--leading expansion.}
\label{fig:kernels}
\vskip-.5cm
\end{figure}
The duality equation~(\ref{dual}) maps the perturbative expansion of
$\gamma$ in powers of $\alpha_s$ at fixed $N$ in the expansion
of
$\chi$ in powers of $\alpha_s$ at fixed $M$ and
conversely. One can thus construct a double leading
expansion~\cite{summing} 
(see
Fig.~\ref{fig:crosses})  which
includes in $\chi$ all terms up to a
given order in the expansion in powers of $\alpha_s$ both at fixed
$M$ and at fixed $\frac{\alpha_s}{M}$.  Its dual $\gamma$  can
be shown to include terms up to the same
order in the expansion in powers of $\alpha$ both at fixed
$\alpha_s$ and at fixed $\frac{\alpha_s}{N}$. 

Using either the
double--leading $\chi$ or the double leading $\gamma$ in the BFKL or
GLAP equation respectively leads to a solution which includes all
terms which are logarithmically enhanced either in $\frac{1}{x}$ or in
$Q^2$ to the given order~\cite{sxres}. 
The result (see Fig.~\ref{fig:kernels}) is close to the GLAP one when
$M\to0$, and close to the BFKL one when $N\to 0$. Because the
perturbative expansion of the BFKL kernel is very poorly behaved, this
resummed result has poor perturbative stability as $N\to 0$.

\subsection{Exchange symmetry}
\label{sec:exchange}
The perturbative instability of the kernel as $N\to0$ can be cured by
observing that the BFKL kernel must be symmetric upon the interchange
$M\to 1-M$, due to the symmetry of the three--gluon vertex
upon the interchange of the radiated and radiating gluon~\cite{ciafresa}.
This symmetry, which is manifest in the LO BFKL kernel, is however
broken beyond the LO of the BFKL expansion by running coupling
correction and by the choice of DIS kinematics~\cite{salam,ciafresa}. Nevertheless, the
symmetry breaking terms can be computed exactly. Hence, one can
symmetrize the double--leading expansion by undoing the symmetry
breaking terms (by changing kinematics and argument of the running
coupling),  then symmetrizing the results, and finally restoring the
original symmetry--breaking kinematics and choice of argument of
$\alpha_s$~\cite{sxsym}.

\begin{figure}[htb]
\begin{center}
\includegraphics[width=\linewidth]{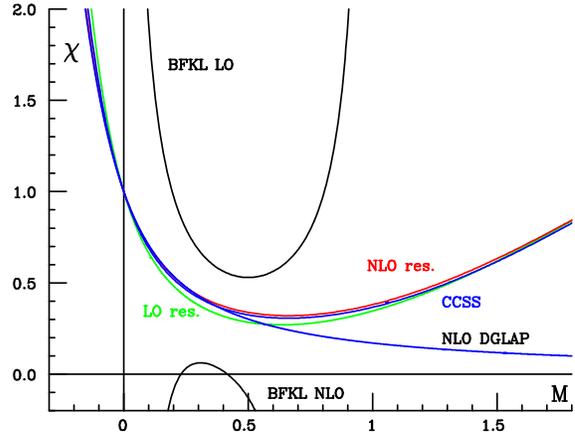} 
\end{center}
\vskip-1.2cm
\caption{The LO and NLO resummed symmetrized double--leading kernels compared
  to the LO and NLO kernels in the BFKL expansion and the NLO GLAP
  kernel. CCSS denotes the corresponding result of Ref.~\cite{ciafresb}
  (from Ref.~\cite{dittmar})
}
\label{fig:symker}
\vskip-.5cm
\end{figure}
The result turns out to be surprisingly stable because of two
features: (a) the anomalous dimension must satisfy the momentum
conservation constraint $\gamma(1,\as)=0$, which by duality implies
$\chi(0,\as)=1$; the anomalous dimension $\gamma(N,\alpha_s)$ 
decreases monotonically as $N$ increases as a consequence of the fact
that a gluon looses momentum when radiating. Combining these two, it follows
that $\chi$ always
has a minimum, because it
must go through the value $\chi=1$ at a pair of values of
$M$ which are   symmetric  about the minimum, located at $M=\frac{1}{2}$. The
minimum is preserved even when the symmetry is broken, in which case
the two ``momentum conservation'' points at which   $\chi=1$ get
shifted to  $M=0$ and  $M=2$.
 One can
further show that when the symmetry breaking is removed, the kernel is
an entire function in the $M$ plane; after restoring the symmetry breaking
it remains free of singularities for ${\rm Re}M>-1$~\cite{sxsym}. 

The perturbative instability of the BFKL kernel
is thus completely removed: see Fig.~\ref{fig:symker}. In this Figure we also
compare results with those obtained by CCSS through a procedure which
is rather different, but shares the following features: (a) all
logarithmically enhanced terms in $Q^2$ and $\frac{1}{x}$ are included
up to NLO (b) the underlying symmetry is implemented up to subleading
terms. The extreme similarity of the results demonstrates the
stability of the procedure.

\subsection{Running coupling}
\label{sec:running}

The double--leading expansion upon which the resummation has been
based so far sums up all terms which are large when $\alpha_s$ is
small, but $\as\xi=\as\ln\frac{1}{x}\sim 1$: namely, a contribution of
the form $F(\as\xi)$ to the splitting function is considered to be of
order $\as^0$. Thus, if the computation is performed at N$^k$L order,
relative 
corrections are of order $\as^{k+1}$
when $\as\to0$ while either $x$ or $\as\xi$ are kept fixed. However, this
does not guarantee that corrections remain small if $\xi\to\infty$
(i.e. $x\to0$) with $\as$ fixed. It is easy to see~\cite{sxap} that any correction
which changes the leading (rightmost in the
$N$ plane)  small $N$ singularity of the anomalous
dimension leads to a contribution to the splitting function
which blows up as $\xi\to\infty$ in comparison to the lower order,
because such a contribution changes the asymptotic $x\to0$
behaviour of the splitting function, and thus of the parton
distribution. 

The leading singularity of  the anomalous dimension is a
simple pole at $N=0$ in the LO (and NLO)  GLAP case. After double leading
resummation  it becomes a square-root 
branch cut: the inverse dual Eq.~(\ref{dual}) of the quadratic
behaviour of the kernel near its minimum. As discussed in
Sect.~\ref{sec:exchange}, the presence of a minimum of the kernel
 is a
generic feature which follows from its symmetry
properties and momentum conservation. The intercept of the kernel
however changes order by order, and it is only its all--order position
which determines the asymptotic small $x$ behaviour. Hence, higher
order corrections to the position of the minimum are asymptotically
large as $\xi\to\infty$:
this suggests
that the all-order location of the minimum of the kernel 
should be treated as a non-perturbative
parameter, to be fitted to the data~\cite{sxres,sxphen}. 

However, running coupling corrections change this state of affairs.
Because the running of the coupling is subleading in
$\ln\frac{1}{x}$, running coupling corrections only affect the
 duality equation~(\ref{dual}) 
through a finite number of terms: at the N$^k$L
logarithmic level, the duality relation is   corrected by a term of
the form
\begin{equation}
\gamma^{(k)}_{\beta_0}=
(\alpha_s\beta_0)^kf^{(k)}_{\beta_0}\left(\frac{\alpha_s}{N}\right),
\label{rccor}
\end{equation}
where the function $f^{(k)}_{\beta_0}\left(\frac{\alpha_s}{N}\right)$
can be calculated
at any given order using suitable operator methods in terms of the
N$^k$LO BFKL kernel~\cite{operator}.

Now, it turns out that whenever the fixed coupling kernel has a
minimum, the running coupling correction $\gamma^{(k)}_{\beta_0}$
is also asymptotically large as $\xi\to\infty$: in fact
 $f^{(k)}_{\beta_0}$  Eq.~(\ref{rccor})
grows as $\as^n\xi^n$ in comparison to the splitting function
computed at LO in the double--leading expansion.
Hence, in order to determine the correct $\xi\to\infty$ limit
we must resum all these terms. This
resummation can be performed exactly  for the
series of terms which grow fastest as $\xi\to\infty$~\cite{sxrun}: the result can
be
 expressed in terms of Airy functions for a
kernel which is linear in $\as$, and Bateman functions for a kernel
with a generic dependence of $\as$~\cite{sxsym}. The (asymptotic, divergent)
expansion of these functions
in powers of $\as$ at fixed $\as\over N$ gives back, to each order in
$\as$,
the
contribution which upon inverse Mellin transform
grows fastest as $\xi\to\infty$ to the 
terms $\gamma^{(k)}_{\beta_0}$ Eq.~(\ref{rccor}).

\begin{figure}[htb]
\begin{center}
\includegraphics[width=.9\linewidth]{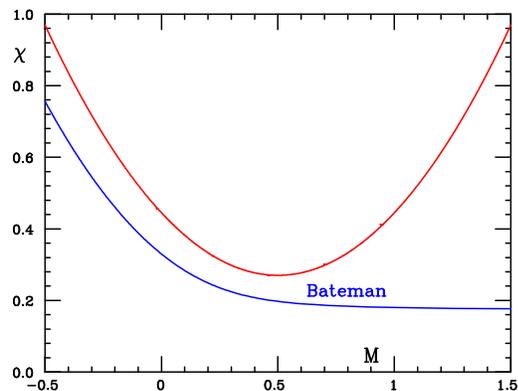} 
\end{center}
\vskip-1.2cm
\caption{Quadratic approximation to the NLO resummed kernel of
  Fig.~\ref{fig:symker}
and its Bateman
  running--coupling resummation.}
\label{fig:bateman}
\vskip-.5cm
\end{figure}
It can then be seen that the resummation of running coupling
corrections changes the nature of the leading singularity: the
fixed-coupling square-root 
branch cut is turned back into
a simple pole. This is shown in Fig.~\ref{fig:bateman}, where we
display the quadratic approximation to the double--leading NLO kernel
of Fig.~\ref{fig:symker}, and its Bateman resummation, i.e. the
anomalous dimension which is obtained from it when running coupling
corrections to Eq.~(\ref{rccor}) it are included to all orders
(computed using 
the dependence on $\alpha_s$ of the NLO resummed result of
Fig.~\ref{fig:symker}).
 
Hence, after running coupling resummation,   the minimum
of the kernel no longer provides the leading small $x$ singularity,
which is instead given by the pole in the Bateman anomalous
dimension. The location and residue of the
Bateman pole are fully determined by the intercept and curvature of the minimum
of the original kernel, and their dependence on $\as$. 

\section{General features of resummed results}
\label{sec:genfeat}

\begin{figure}[htb]
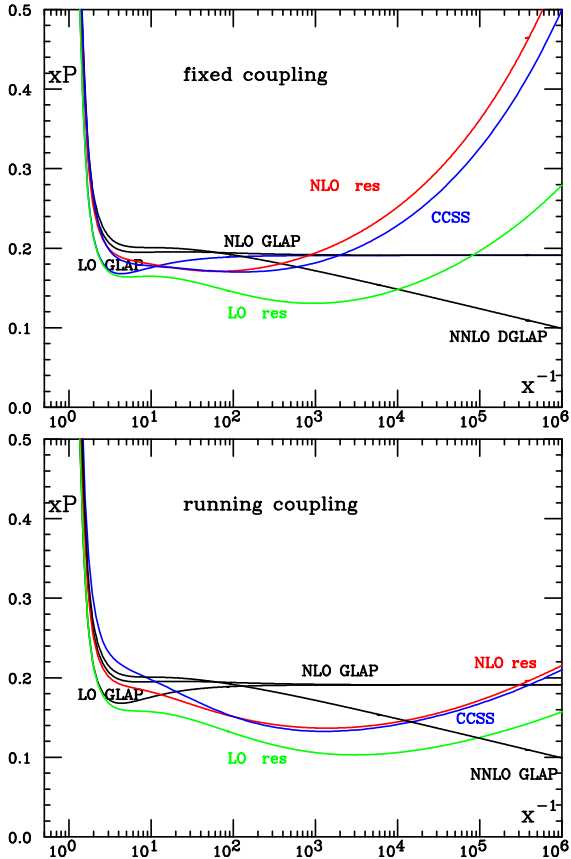

\begin{center}
\includegraphics[width=\linewidth]{abfpggb0.ps}\\ 
\includegraphics[width=\linewidth]{abffinalp.ps} 
\end{center}
\vskip-1.2cm
\caption{The resummed splitting function with $n_f=0$ and
  $\alpha_s=0.12$ with fixed coupling (top) and running coupling
  (bottom).}
\vskip-.7cm
\label{fig:finsplnfzero}
\end{figure}
The crucial feature of the resummation procedure summarized in the
previous section is that at each step the contributions which are
included in the anomalous dimension on top of its standard GLAP
fixed--order expansion in powers of $\as$ at fixed $N$ can be expanded
out perturbatively, so that it is possible to obtain a fully resummed
expression by simply combining all contributions, and subtracting the
double counting terms. The procedure can be performed order by order in
perturbation theory, by starting with the double--leading expansion
of Fig.~\ref{fig:crosses} to any given order, and then improving it as
discussed in the previous section.

The resummed anomalous dimension has
then schematically the form
\begin{eqnarray}
&&\gamma^{rc}_{\Sigma\,NLO}(\as(t),N)
=\gamma^{rc,\,pert}_{\Sigma\,NLO}(\as(t),N)\nonumber\\&&+
\gamma^B(\as(t),N)- 
\gamma^B_s(\as(t),N)\nonumber \\&&
-\gamma^B_{ss}(\as(t),N)
-\gamma^B_{ss,0}(\as(t),N)\\&&+\gamma_{\rm match}(\as(t),N)
+\gamma_{\rm mom}(\as(t),N),\nonumber
\label{resad}
\end{eqnarray}
where \begin{itemize}
\item $\gamma^{rc,\,pert}_{\Sigma\,NLO}(\as(t),N)$ is the fixed--coupling
resummed anomalous dimension displayed in Fig.~\ref{fig:symker},
obtained by duality Eq.~(\ref{dual}) from the kernel which in turn is
found by symmetrization (Sect.~\ref{sec:exchange}) of the
NLO double--leading kernel (Figs.~\ref{fig:crosses},\ref{fig:kernels});
\item $\gamma^B(\as(t),N)$  is the Bateman anomalous dimension Fig.~\ref{fig:bateman};
\item $\gamma^B_s(\as(t),N)$, 
$\gamma^B_{ss}(\as(t),N)$ $\gamma^B_{ss,0}(\as(t),N)$ are double
counting subtractions between the previous two, namely the
contributions to the LO and NLO terms
$\gamma^{(k)}_{\beta_0}$ Eq.~(\ref{rccor}) which grow fastest as $\xi\to\infty$;
\item $\gamma_{\rm mom}$ subtracts subleading terms which would
  otherwise spoil
  momentum conservation;
\item $\gamma_{\rm match}$ subtracts any contribution which deviates
  from NLO GLAP and at large $N$ (which corresponds to large $x$) doesn't drop at least as $\frac{1}{N}$.
\end{itemize}
Note that the last two contributions are formally subleading: they are
included in order to improve the matching to the GLAP anomalous
dimension, in that they remove from the resummation subleading
contributions which may be non-negligible in the large $x$ region
where the resummation is not supposed to have any effect.

In Figure~\ref{fig:finsplnfzero} we display the 
splitting functions obtained from Mellin inversion of the resummed
anomalous dimension Eq.~(\ref{resad}) at the fixed coupling level and at the
running coupling level --- i.e. respectively without and with 
the inclusion of the Bateman
contribution 
$\gamma^B(\as(t),N)$ and its associate double counting subtractions.
The result is also compared to the CCSS result Ref.~\cite{ciafresb} (from
Ref.~\cite{dittmar}).
The resummed expansion is seen to be stable (the LO and NLO results are
close), all the more so at
the running coupling level. The resummed result matches smoothly to
the GLAP result in the large $x\gsim 0.1$ region, but at small $x$ it
is free of the instability which the GLAP expansion shows already at
NNLO. 

The comparison between results obtained with  the ABF method discussed
here and those by CCSS~\cite{ciafresb} 
is illuminating in various
respects. The CCSS approach also includes the three ingredients discussed in
the previous section  --- double--leading resummation, symmetrization of
the kernel, and running coupling corrections --- 
but the implementation is rather different. In particular, the
resummation of running
coupling corrections is obtained by numerical solution of the
running--coupling BFKL equation~(\ref{bfkl}) (see
Ref.~\cite{dittmar}). The closeness of results obtained by CCSS and
ABF at the fixed coupling level reflects the closeness of the kernels
Fig.~\ref{fig:symker}. The fact that CCSS and ABF results are even closer at the
running coupling levels follows from the softening of
resummation effects due to  asymptotic 
freedom. Also, the fact that exact numerical resolution of the running--coupling
BFKL equation (in the CCSS approach) followed by a numerical
extraction of the associate anomalous dimension, and the resummation of the
leading running coupling corrections Eq.~(\ref{rccor}) in terms of a
Bateman function (in the ABF approach) lead to a result which manifestly coincides for all
$x\lsim 0.03$ supports the accuracy of both procedures. 

It is
interesting to observe however that the CCSS resummed result
shows a significant deviation from the unresummed GLAP result even for
$x\gsim 0.1$, which is not seen in the ABF result. Because this
effect  is only present at the running coupling level, it is likely to
be due to the fact that running coupling terms lead to contributions
which survive in the large $x$ region, where these terms are formally
subleading (recall that the running coupling contributions were
selected on the basis of their behaviour as $\xi\to\infty$
i.e. $x\to0$). In the ABF approach (but not in the CCSS approach)
these contributions are subtracted
through the term $\gamma_{\rm match}(\as(t),N)$ in
Eq.~(\ref{resad}). We have found this subtraction to be necessary in order for the
resummed results not to differ from the fixed--order ones in the
$x\gsim 0.1$ region.

\begin{figure}[htb]
\begin{center}
\includegraphics[width=.8\linewidth]{abfintercepts.ps}\\
\vskip.3cm
\hskip-.5cm\includegraphics[width=\linewidth]{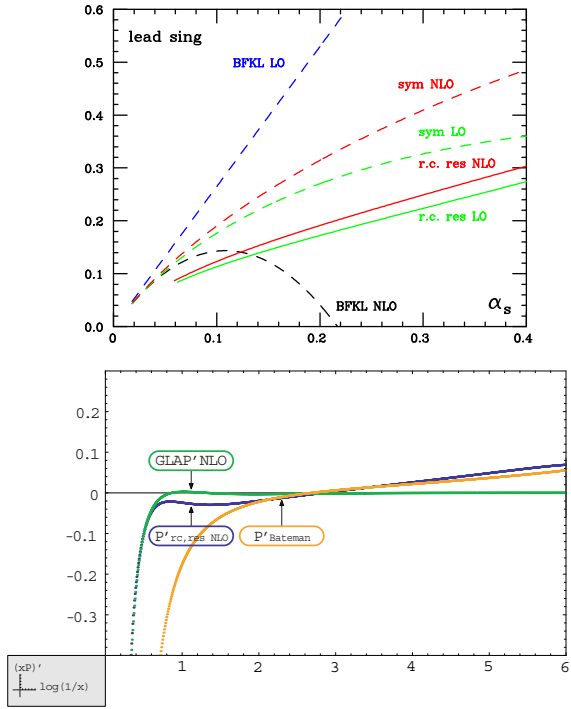} 
\end{center}
\vskip-1.2cm
\caption{The leading small $x$ singularity vs. $\alpha_s$ (top) and
  the slope of the splitting function vs. $\xi=\ln\frac{1}{x}$ 
(bottom, from Ref.~\cite{frugiuele}).}
\label{fig:intercepts}
\vskip-.7cm
\end{figure}
More detailed insight into the generic features of the resummation can be
obtained by studying its small $x$ behaviour. As already mentioned,
this is largely determined by the position of the leading (rightmost)
singularity of the anomalous dimension in the $N$ plane. 
For the LO and NLO GLAP anomalous dimension this is
located at $N=0$, while at the resummed level it 
is plotted as a function of $\alpha_s$ in
Fig.~\ref{fig:intercepts}: the large and perturbatively unstable
corrections at the BFKL level (compare Fig.~\ref{fig:kernels}) 
are turned into more moderate and stable correction (compare
Fig.~\ref{fig:symker}) by the symmetrization, and further reduced and
stabilized by the running of the coupling. 

The relative importance of
various contributions to the resummation is elucidated by comparing
in Fig.~\ref{fig:intercepts}
the slope of the resummed running--coupling
splitting function (Fig.~\ref{fig:finsplnfzero}) with respect to $\xi$
 to that of the NLO GLAP result, and that of
the splitting function obtained by inverse Mellin transform of the
Bateman anomalous dimension $\gamma^B(\as(t),N)$ in
Eq.~(\ref{resad}). It is clear that for $x\gsim0.1$ the slope (and
thus, up to a constant, the splitting function) is
determined by the GLAP result, and for $x\lsim 10^{-2}$ by the Bateman
result~\cite{frugiuele}. Note that this is despite the fact that it is
only for tiny values of
$x\lsim10^{-15}$ that the Bateman asymptotic slope reduces (to within
about 10\%) to  its
asymptotic value $N_0$ (the location of the leading
singularity)~\cite{frugiuele}. 

Hence the qualitative features of the resummation
are essentially determined by matching to the GLAP result
the Bateman anomalous dimension,
which in turn is fully determined by resummation of the running
coupling terms Eq.~(\ref{rccor}), and thus by the intercept and shape of the
minimum of the symmetrized kernel displayed in
Fig.~\ref{fig:symker}~\cite{sxsym}. The insensitivity of these to the details of the
resummation procedure explains the stability of the results. The
dominant feature of the result in the HERA region $10^{-2}\gsim x\gsim
10^{-4}$ is  a dip in the splitting function which
results from this
matching (see Ref.~\cite{ciafdip} for an alternative discussion of
this feature of resummed results).

A resummed gluon splitting function was also presented in
Refs.~\cite{T1,TW}. The agreement is reasonable at the qualitative level,
but the resummed splitting function appears to display a stronger rise
at small $x$ and a somewhat more pronounced dip at intermediate
$x$. This may be due to the fact that in Ref.~\cite{T1,TW} 
no symmetrization of the kernel is
performed: this, as discussed above, leads to resummed results which
tend to be unstable upon perturbative corrections as $x\to0$. 

\section{Quarks and phenomenology}
\label{sec:gtoq}

Going from a resummation of the evolution equations at $N_f=0$  to fully resummed
physical observables requires two ingredients: the inclusion of the
quark sector in the resummation, and resummed partonic cross
sections. 

\subsection{The resummed splitting function matrix}
\label{sec:factsch}

An extension of small $x$ BFKL-like evolution equations to
the quark sector has been attempted in Ref.~\cite{ciafquarks}, but
results in a fully consistent factorization scheme are still not
available.
However, it turns out that this construction is not necessary for the
determination
of the matrix of resummed splitting functions. This is
due to the fact that the resummation only
affects one of the two eigenvectors of the singlet anomalous dimension
matrix. Therefore, in order to obtain coupled resummed evolution
equations for singlet quarks and gluons it is sufficient to fix the
factorization scheme at the resummed level~\cite{symphen}.

Indeed, calling $\gamma^+$ and $\gamma^-$ the two eigenvectors of the
anomalous dimension matrix, if only  
contributions to $\gamma$ which are singular as $N\to0$ are included, $\gamma^-=0$
and thus
\begin{eqnarray}
&&\gamma^{gg}+\gamma^{qq}=\gamma^+;\nonumber\\
&&\gamma^{gg}\gamma^{qq}=\gamma^{qg}\gamma^{gq}.\label{scheme}
\end{eqnarray}
 up to
    nonsingular terms.
Furthermore, in the $\overline{\rm MS}$ and related schemes,
$\gamma^{gq}=\frac{C_F}{C_A}\gamma^{gg}$ (up to
    nonsingular terms)~\cite{ch}. Hence, combining the determination of
 $\gamma^+$ with the knowledge of $\gamma^{qg}$ in the
    $\overline{\rm MS}$ scheme, which was computed at the resummed level
    in Ref.~\cite{ch}, the anomalous dimension
    matrix is fully determined.

\begin{figure}[htb]
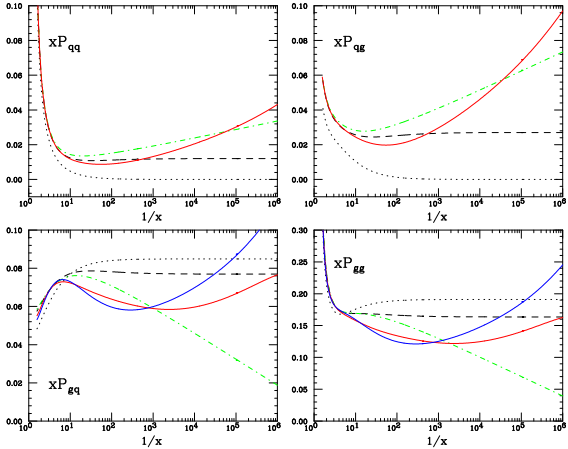

\begin{center}
\includegraphics[width=.49\linewidth]{abfpqq.ps}
\includegraphics[width=.49\linewidth]{abfpqg.ps}\\
\includegraphics[width=.49\linewidth]{abfpgq.ps}
\includegraphics[width=.49\linewidth]{abfpgg.ps}\\
\end{center}
\vskip-1.2cm
\caption{The splitting function matrix with $n_f=4$,
  $\alpha_s=0.2$. The curves correspond to LO (dotted), NLO (dashed),
  NNLO (dot-dash), resummed (solid). The two different resummed curves
  (in the gluon sector) correspond to the $\overline{\rm
    MS}$ (steeper at small $x$) and Q$_0\overline{\rm
    MS}$ factorization schemes.}
\label{fig:splfct}
\vskip-.7cm
\end{figure}The actual computation of the full splitting function
matrix with $n_f\not=0$
 entails some technical complications, which have
only recently been solved in Ref.~\cite{symphen}.
Firstly, when $n_f\not=0$, the eigenvectors of the anomalous dimension
  matrix develop an unphysical singularity at the value of $N$ where
  the two eigenvectors cross. This singularity cancels in the solution
  to evolution equations, and the cancellation must be enforced
  exactly throughout the double--leading resummation and
  symmetrization: if it were spoiled by subleading terms, this would
  lead to a spurious small $x$ rise of parton distributions. 
Furthermore, the quark--sector anomalous dimension $\gamma_{qg}$  was determined
  in Ref.~\cite{ch} in the $\overline{\rm MS}$ scheme, while the
  double--leading resummation is most naturally performed in the Q$_0$
  scheme~\cite{qz,schemes}: in the  $\overline{\rm MS}$ scheme the
  running coupling terms Eq.~(\ref{rccor}) are 
  factored order by order in the
  coefficient function. Because their resummation determines the
  leading small $x$  behaviour, as discussed in
  Sect.~\ref{sec:running}, it is more convenient to perform the
  resummation in a scheme where they are included in the splitting
  function. Whereas the scheme change from the Q$_0$ scheme to
  $\overline{\rm MS}$   in the
  pure--gluon case
  was worked out previously~\cite{qz,schemes}, its construction in the
  presence of quarks is nontrivial~\cite{summing,symphen}.

The resummed matrix of splitting functions in  the $\overline{\rm MS}$
and Q$_0\overline{\rm MS}$ factorization schemes
is compared to the unresummed result in
Fig.~\ref{fig:splfct}. Whereas a detailed comparison with the CCSS
splitting function matrix~\cite{ciafquarks} is not possible, because
CCSS results are given in a scheme which differs from the standard
$\overline{\rm MS}$ scheme by an amount which is  undetermined
beyond fixed NLO, a qualitative comparison shows reasonable
agreement. 

Quark sector splitting functions have also been given in
Ref.~\cite{TW} (see also Ref.~\cite{heralhc08}). 
Their agreement with those of
Fig.~\ref{fig:splfct} is not so good: the $P_{qg}$
splitting function shows a much stronger small $x$ rise, and a sizable
deviation from the NLO result at large $x\gsim 0.1$. The latter
feature can be understood as a consequence of the fact that in
Refs.~\cite{T1,TW} no matching to large $x$ is performed:
contributions from the small $x$ resummation in the large $x$ region
are not subtracted. The former feature is likely to be due to the fact
that results in Ref.~\cite{TW} are not determined in a fully
consistent factorization scheme. In fact, the resummation is performed
in the Q$_0$ scheme, but it is then combined with
$\overline{\rm MS}$ (or DIS) scheme coefficient functions: 
in the TW approach, the issue of
the scheme transformation from Q$_0$ to $\overline{\rm MS}$  is still
unresolved. Because of the aforementioned interplay between the scheme
choice and the resummation of running coupling singularity, this
inconsistency is
likely to affect strongly the small $x$ behaviour.

\subsection{Coefficient function resummation}
\label{sec:cfres}

The leading small $x$ contributions to partonic cross sections
are known to all orders for a small but
increasing number of physical processes: they were first computed
for heavy quark photo--and electroproduction in Ref.~\cite{hq} (later
extended to hadroproduction in Ref.~\cite{be}), they have been
determined for deep-inelastic
scattering in Ref.~\cite{ch} and more recently for Higgs
production~\cite{higgs} and the Drell-Yan processes~\cite{dy}.

\begin{figure}[htb]
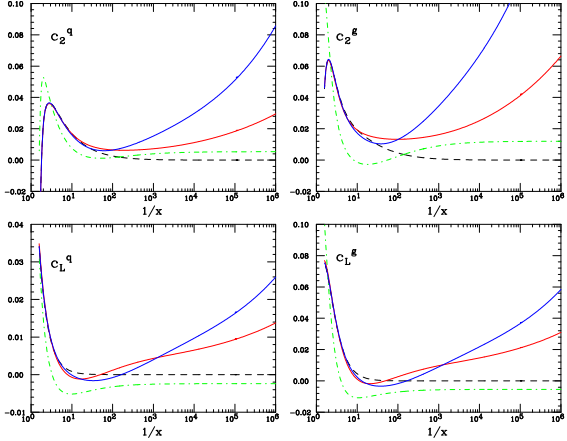

\begin{center}
\includegraphics[width=.49\linewidth]{abfc2q.ps}
\includegraphics[width=.49\linewidth]{abfc2g.ps}\\
\includegraphics[width=.49\linewidth]{abfclq.ps}
\includegraphics[width=.49\linewidth]{abfclg.ps}\\
\end{center}
\vskip-1.2cm
\caption{The matrix of coefficient functions with $n_f=4$,
  $\alpha_s=0.2$. The curves correspond to NLO (dashed),
  NNLO (dot-dash), resummed (solid). The two different resummed curves
  (in the gluon sector) correspond to the $\overline{\rm
    MS}$ (steeper at small $x$) and Q$_0\overline{\rm
    MS}$ factorization schemes.}
\label{fig:cofun}
\vskip-.7cm
\end{figure}
Expressions for coefficient functions in the NLO of the
double--leading expansion were already constructed in
Ref.~\cite{sxphen}, where, however, the running coupling terms
discussed in Sect.~\ref{sec:running} were still left unresummed, thereby
simplifying issues of scheme dependence, but at the cost of not
reproducing the correct small $x$ behaviour. However, running
coupling corrections to the resummation of coefficient functions also
grow as $\xi\to\infty$, analogously to the running coupling
corrections to splitting functions Eq.~\ref{rccor}: they must be 
resummed lest physical observables develop unphysical
singularities which leads to a spurious small $x$
growth~\cite{cfresum}. 

The resummation can be performed exactly~\cite{cfresum} when
the double--leading expression of the coefficient function is known in
closed form. This is however not the case for the $F_2$ deep-inelastic
coefficient function in the $\overline{\rm MS}$ scheme, for which only
a series expansion generated by a recursion relation is
available~\cite{ch}. In such case, the dominant running coupling corrections
to the coefficient function can be resummed through a divergent asymptotic
expansion, which may be summed by the Borel method~\cite{symphen}. The
ensuing resummed coefficient functions are displayed in
Fig.~\ref{fig:cofun}. Resummed coefficient functions were also
presented in Ref.~\cite{TW} (also including running coupling
resummation) and are qualitatively similar, though a detailed
comparison is hampered by the fact that  the factorization
scheme used there is different (DIS instead of $\overline{\rm MS}$).

\subsection{Parton distributions and structure functions}
\label{sec:strfn}
\begin{figure}[htb]
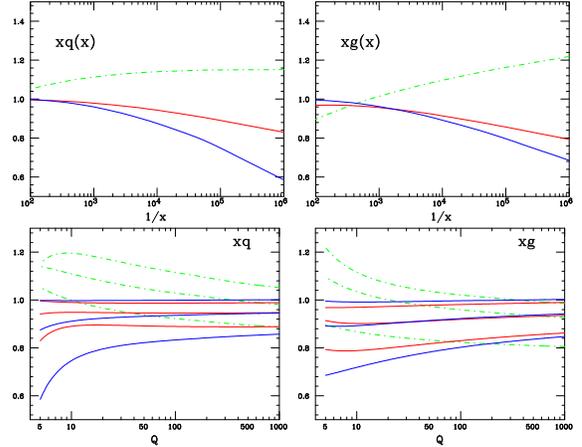

\begin{center}
\includegraphics[width=.49\linewidth]{abfqbcx.ps}
\includegraphics[width=.49\linewidth]{abfgbcx.ps}\\
\includegraphics[width=.49\linewidth]{abfqbc.ps}
\includegraphics[width=.49\linewidth]{abfgbc.ps} 
\end{center}
\vskip-1.2cm
\caption{The ratio of the resummed (solid) or NNLO (dot-dashed) to NLO
singlet quark and gluon distributions as a function of $x$ at the
 scale $Q_0=5$~GeV (top) and as a function of $Q$  at fixed
 $x=10^{-2},\>10^{-4},\>x=10^{-6}$ (bottom; the smallest $x$ is the
 lowest curve in the resummed case and
 the highest  at NNLO). The ratios  are determined assuming
that the structure  functions $F_2$ and $F_L$ are kept fixed at the
 scale $Q_0=5$~GeV. The two different resummed curves
  correspond to the $\overline{\rm
    MS}$ (smaller at small $x$) and Q$_0\overline{\rm
    MS}$ factorization schemes. 
}
\label{fig:pdfbc}
\vskip-.7cm
\end{figure}
Combining  the ingredients discussed so far it is
possible to determine resummed predictions for deep--inelastic
structure functions. Eventually, these should be used, together with
resummed expressions for other physical processes, for the
determination of parton distributions at the resummed level.

An estimate of the impact of resummation on parton distributions can
be obtained by first computing the structure functions $F_2$ and $F_L$
with some typical ``toy'' set of NLO parton distributions (PDFs), and then 
assuming that the structure functions 
are kept fixed at some scale: this is then enough to determine the resummed
singlet quark and gluon distribution at that scale. The effect on
PDFs is close to that which would be obtained if PDFs were determined
from a fit to
DIS data mostly clustered around  that scale. Results for the typical HERA
(compare Fig.~\ref{fig:kinreg}) scale choice of Q$^2=25$~GeV$^2$
are shown in Fig.~\ref{fig:pdfbc}, where we display the
singlet quark and gluon distributions as a function of $x$ at this
starting scale, and then as a function of $Q^2$ for various $x$
values. Results are shown as a ratio of the resummed or NNLO result
to the NLO one. The impact of the resummation at the ``HERA scale'' 
is comparable to that of
NNLO corrections, but it goes in the opposite direction: it tends to
suppress the starting PDFs while the NNLO tends to
enhance them. When evolving up the
differences tend to be washed out
because of asymptotic freedom.

The PDFs displayed in Fig.~\ref{fig:pdfbc} can then
be used to determine $K$-factors for  resummed structure functions:
these were computed in Ref.~\cite{symphen}. There  the 
factorization
and renormalization scale dependence of the of results were also studied, and
shown to be small. Indeed, once all ingredients are included, 
the resummed perturbative expansion of physical
observables turns out to be as good as the standard fixed--order expansion.

\begin{figure}[htb]
\begin{center}
\includegraphics[width=\linewidth]{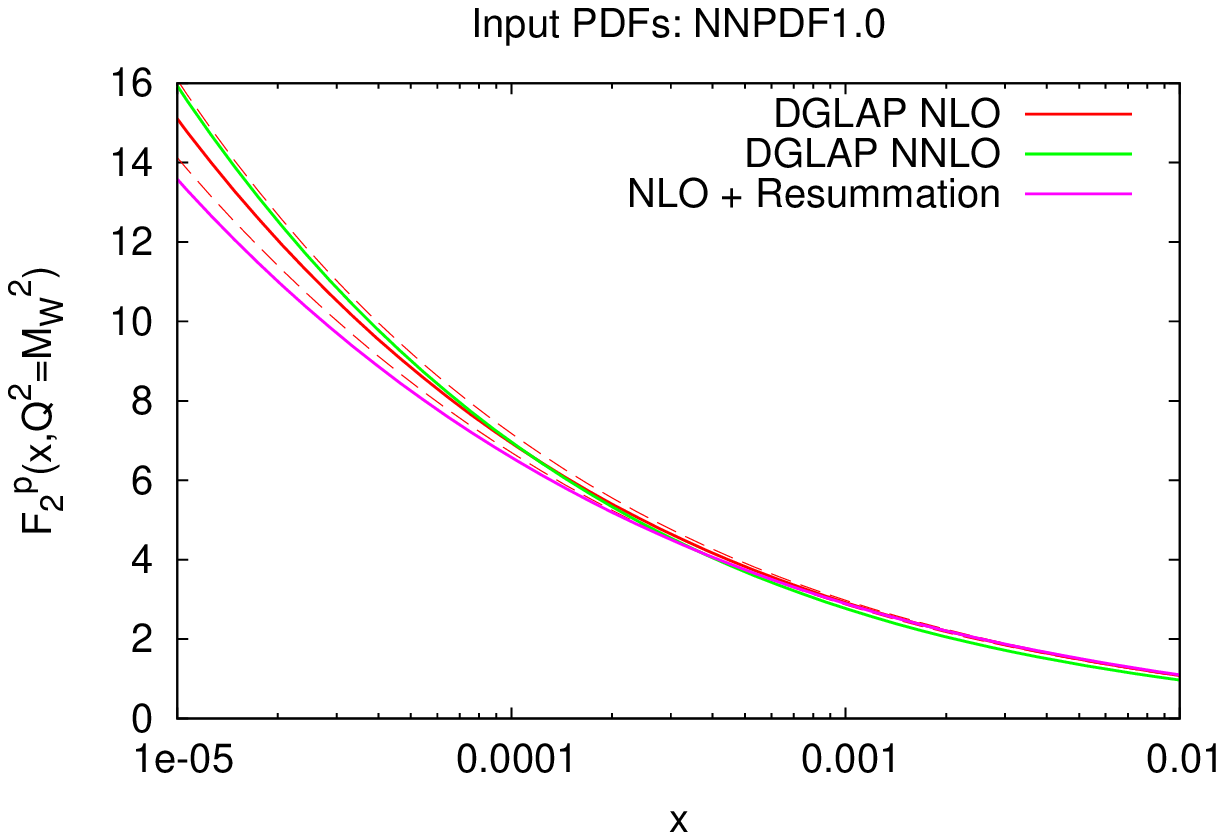}\\
\includegraphics[width=\linewidth]{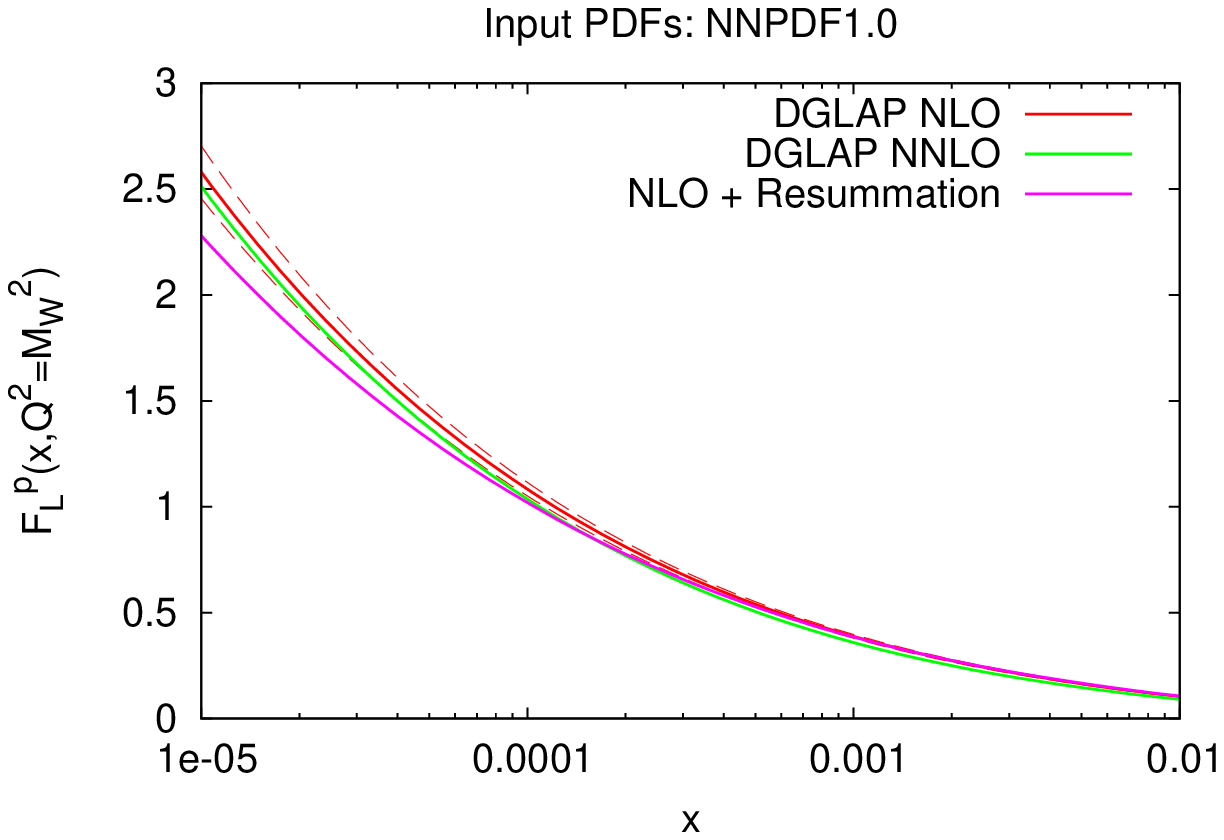} 
\end{center}
\vskip-1.2cm
\caption{The deep--inelastic structure functions $F_2$  and $F_L$
  at $Q^2=M_W^2$, computed at NLO, NNLO and resummed level. The parton
  distributions are from the NLO
  NNPDF1.0 set in the NLO case, while at the NNLO and resummed level
  they are computed using $K$--factors as discussed in the text. The
  PDF uncertainty band is shown for the NLO prediction, while only the
  central prediction is displayed in the other cases.}
\label{fig:strfctn}
\vskip-.7cm
\end{figure}
The dominant
qualitative feature of the $K$--factors is that resummation leads to
a suppression of the structure functions $F_2$ and $F_L$ at small
$x$. The $K$--factors were compared to those determined using
 the TW approach~\cite{TW} in Ref.~\cite{heralhc08}. The main
 differences are related to the fact that TW $K$-factors can differ
 sizably from one even at $x\gsim 10^{-2}$ where the ABF result matches
   smoothly to the NLO one, and that the TW $K$-factors  
still show a marked scale
   dependence at large $Q^2\gsim10^{4}$~GeV$^2$ where the scale
   dependence of the ABF result flattens out completely. These
   differences are likely to be due to the various features of the TW
   approach mentioned above: in particular, the lack of
   matching terms to large $x$, and the fact that the factorization
   scheme is not treated in a fully consistent way. Indeed, there
   appears~\cite{symphen} to be a significant cancellation of
 scheme dependence between evolved  parton distributions and coefficient
 functions, which is inevitable spoiled if these are not determined in
 the same factorization scheme.

$K$--factors can also be used to determine
resummed predictions in an approximate way, by first, computing the
$K$ factors as discussed above, but using a  realistic underlying set
of PDFs, then verifying that the values of the $K$ factors are
reasonably independent of the choice of underlying PDFs, 
and finally  using the $K$--factors to correct NLO
predictions obtained from any PDF set. 

In Fig.~\ref{fig:strfctn} we
show the 
structure functions $F_2$ and $F_L$ at the scale $Q^2=M_W^2$ obtained
applying either resummed or NNLO $K$--factors to the NLO prediction
from the NNPDF1.0 parton set~\cite{nnpdf}. Interestingly, the
suppression due to the resummation in the small $x\lsim10^{-4}$ region is
larger than the one--sigma band due to PDF uncertainties. 

These results show that resummation is relevant
 for deep--inelastic scattering at a high--energy
hadron-electron collider such as a collider
based on the LHC (LHeC~\cite{lhec}).
It will  be interesting to study  the impact on
precision LHC processes such as W and Z production of resummation
corrections, both from their direct effect on partonic cross
sections, and their indirect effect due to their impact on the
extraction of parton distributions. 

In summary, we have seen that the impact of
resummation is 
as large as that of NNLO corrections in the HERA region. The
theoretical framework on which resummed results are based is now on a
similar footing as that of standard fixed order perturbation theory,
whose reliability is thus extended also to kinematical regions,
relevant for HERA and the LHC, where NNLO results become unstable.

{\bf Acknowledgements:} We thank M.~Ciafaloni, D.~Colferai, G.~Salam,
A.~Stasto, R.~Thorne and C.~White for various discussions, and J.~Rojo
for help with the NNPDF prediction of Fig.~\ref{fig:strfctn}. This work
was partly supported by the Scottish Universities' Physics Alliance, by the
Italian PRIN program for 2007-08, and by the Marie Curie
Research and Training network HEPTOOLS under contract
MRTN-CT-2006-035505.

\end{document}